\documentclass[12pt]{iopart}

\usepackage{iopams}  
\usepackage{graphicx}
\usepackage{subfig}
\usepackage{xcolor}
\usepackage{ulem}
\usepackage{lscape}

\newcommand{\bq}{\begin{equation}}
\newcommand{\eq}{\end{equation}}
\newcommand{\bqn}{\begin{eqnarray}}
\newcommand{\eqn}{\end{eqnarray}}
\newcommand{\nb}{\nonumber}
\newcommand{\lb}{\label}

\begin{document}

\title{Spherically Symmetric Analytic Solutions and Naked Singularities in Einstein-Aether Theory}

\author{R. Chan$^{1}$, M. F. A. da Silva$^{2}$ and V. H. Satheeshkumar$^{3}$} 

\address{
			$^{1}$Coordena\c{c}\~{a}o de Astronomia e Astrof\'{i}sica,
			Observat\'{o}rio Nacional (ON), Rio de Janeiro, RJ 20921-400, Brazil
			\\
			$^{2}$Departamento de F\'{i}sica Te\'{o}rica, 
			Universidade do Estado do Rio de Janeiro (UERJ), Rio de Janeiro, RJ 20550-900, Brazil
			\\
			$^{3}$Departamento de F\'{\i}sica, Universidade Federal do Estado do Rio de Janeiro (UNIRIO), Rio de Janeiro, RJ 22290-240, Brazil
			}
			
\ead{chan@on.br, mfasnic@gmail.com, vhsatheeshkumar@gmail.com}

\begin{abstract}
In the present work we analyze all the possible spherically symmetric exterior vacuum solutions allowed by the Einstein-Aether theory with static aether. We show that there are four classes of solutions corresponding to different values of a combination of the free parameters, $c_{14}=c_1+c_4$, which are: $ 0 < c_{14}<2$, $c_{14} < 0$,  $c_{14}=2$ and $c_{14}=0$. We present explicit analytical solutions for $c_{14}=3/2, 16/9, 48/25, -16, 2$ and $0$. The first case has some pathological behavior, while the rest have all singularities at $r=0$ and are asymptotically flat spacetimes. For the solutions $c_{14}=16/9, 48/25\, \mathrm{\, and \,}\, -16$ we show that there exist no horizons, neither Killing nor universal horizon, thus we have naked singularities. Finally, the solution for $c_{14}=2$ has a metric component as an arbitrary function of radial coordinate, when it is chosen to be the same as in the Schwarzschild case, we have a  physical singularity at finite radius, besides the one at $r=0$. This characteristic is completely different from General Relativity.
\end{abstract}

\section{Introduction}

The Lorentz invariance is an exact symmetry of special relativity, quantum field theories and the standard model of particle physics, and also a local symmetry in freely falling inertial frames in General Relativity \cite{Moore:2013sra}. The Lorentz violation in matter interactions is highly constrained by several precision experiments, see \cite{Bars:2019lek} for the latest example, while similar studies in gravity are not as well explored. With this motivation, Jacobson and collaborators introduced and analyzed a general class of vector-tensor theories called the Einstein-Aether (EA) theory \cite{Jacobson:2000xp} \cite{Eling:2003rd} \cite{Jacobson:2004ts} \cite{Eling:2004dk} \cite{Foster:2005dk}.

The first paper to investigate spherical static vacuum solutions in the EA theory was presented by Eling and Jacobson in 2006 \cite{Eling2006}. In that paper, the authors found a family of analytical solutions for the metric functions, up to the inversion of a
transcendental function,  assuming an aether vector proportional to the timelike Killing field. These solutions do not depend on the parameters $c_2$ and $c_3$ of the EA theory  but only depend on the combination $ c_1 + c_4 $, which for simplicity we denote as $c_{14}$. such that $ c_1 + c_4 < 2$. For $ c_1 + c_4 > 2$  the coupling constant $G$ becomes negative, implying that the gravity is repulsive, and for $ c_1 + c_4 =0$ this one is exactly the Newtonian gravitational constant. The authors have shown when $c_1+c_4 = 0$ the Schwarzschild solution can be obtained and also explored some solutions for different choices of  $0 \le c_1+c_4 < 2$. In another paper, Eling, Jacobson and Miller \cite{Eling2007} studied a perfect fluid, in order to model a neutron star, and consider the vacuum solution given in the previous paper \cite{Eling2006} for any $ c_1 + c_4 <2$.  Almost the complete literature on black holes in EA theory can be found in the papers \cite{Foster2006}-\cite{Zhang:2020too}.

Since this article deals with naked singularities, some introductory remarks are in order. In 1965, Roger Penrose proved the first modern singularity theorem \cite{Penrose:1964wq} for which he is recognized with a Nobel Prize in physics \textcolor{black}{last} year \cite{Nobel2020}. In this paper, Penrose introduced the notion of closed trapped surface for the first time to prove the gravitational collapse in GR inevitably leads to singularities. The most important physical consequence of the singularity theorem is to know whether these singularities can be observed. Typically the singularities that arise in the solutions of Einstein field equations are hidden within event horizons. The singularities that are not hidden inside the event horizon are called naked singularities. The Cosmic Censorship Conjecture (CCC), introduced by Penrose in 1969 \cite{Penrose:1969pc}, does not permit singularities void of an event horizon as the end state of gravitational collapse \textcolor{black}{for generic regular initial data and for a suitable matter system}.  Although there is no satisfactory mathematical formulation or the proof of CCC,  there are many examples of dynamical collapse models which lead to a black hole or a naked singularity as the collapse end state, depending on the nature of the initial data (see e.g. \cite{rev1}-\cite{rev8} and references therein). Two major themes in this context are either mathematically prove or disprove the CCC or provide counter examples. We demonstrate here that EA theory provides a natural counter example without needing us introduce any exotic matter. Naked singularities are of current interest because they have observational properties quite different from a black hole. Besides, theoretically these regions of extreme gravity might have some hints of quantum gravity. If observed, they will provide exciting laboratory for the discovery of new physics. Most up-to-date account on naked singularities can be found in \cite{Joshi:2015uoq}. Some of the observational consequences of naked singularities explored recently can be found in the references \cite{Joshi2014} -- \cite{Shaikh2019}. 

The paper is organized as follows.  The Section $2$ briefly outlines the EA theory, whose field equations are solved for a general spherically symmetric static metric in Section $3$. In Sections $4$, $5$ and $6$ we present the explicit analytical solutions. We summarize our results in Section $7$.

\section{Field equations in the EA theory }

The general action of the EA theory is given by  
\bq 
S =  \frac{1}{16\pi G}\int \sqrt{-g}~(R+L_{\rm aether}+L_{\rm matter}) d^{4}x,
\label{action}
\eq
where, the first term defined by $R$ is the usual Ricci scalar, and $G$ the EA coupling constant.
The second term, the aether Lagrangian is given by
\bq 
L_{\rm aether} =  [-K^{ab}{}_{mn} \nabla_a u^m
\nabla_b u^n +
\lambda(g_{ab}u^a u^b + 1)],
\lb{LEAG}
\eq
where the tensor ${K^{ab}}_{mn}$ is defined as
\bq 
{K^{ab}}_{mn} = c_1 g^{ab}g_{mn}+c_2\delta^{a}_{m} \delta^{b}_{n}
+c_3\delta^{a}_{n}\delta^{b}_{m}-c_4u^a u^b g_{mn},
\lb{Kab}
\eq
being the $c_i$ dimensionless coupling constants, and $\lambda$
a Lagrange multiplier enforcing the unit timelike constraint on the aether, and 
\bq
\delta^a_m \delta^b_n =g^{a\alpha}g_{\alpha m} g^{b\beta}g_{\beta n}.
\eq
Finally, the last term, $L_{\rm matter}$ is the matter Lagrangian, which depends on the metric tensor and the matter field.

In the weak-field, slow-motion limit EA theory reduces to Newtonian gravity with a value of  Newton's constant $G_{\rm N}$ related to the EA coupling constant $G$ in the action (\ref{action}) by \cite{Garfinkle2007},
\bq
G = G_N\left(1-\frac{c_{14}}{2}\right).
\lb{Ge}
\eq
Here, the constant $c_{14}$ is defined as
\bq
c_{14}=c_1+c_4.
\lb{beta}
\eq
Note that if $ c_{14} =0$ the EA coupling constant $G$ becomes the Newtonian coupling constant $G_N$, without necessarily imposing $c_1=c_4=0$. For $ c_{14} > 2$  the coupling constant $G$ becomes negative, implying that the gravity is repulsive. The coupling constant vanishes when $c_{14} = 2$. Thus, physically interesting region is {$c_{14} \le 2$} where the Newtonian limit can be recovered.

The field equations are obtained by extremizing the action with respect to independent  variables of the system. The variation with respect to the Lagrange multiplier $\lambda$ imposes the condition that $u^a$ is a unit timelike vector, thus 
\bq
g_{ab}u^a u^b = -1,
\label{LagMul}
\eq
while the variation of the action with respect $u^a$, leads to \cite{Garfinkle2007}
\bq
\nabla_a J^a\;_b + c_4 a_a \nabla_b u^a + \lambda u_b = 0,
\eq
where,
\bq
J^a\;_m=K^{ab}\;_{mn} \nabla_b u^n,
\eq
and
\bq
a_a=u^b \nabla_b u_a.
\eq
The variation of the action with respect to the metric $g_{mn}$ gives the dynamical equations,
\bq
G^{Einstein}_{ab} = T^{aether}_{ab} +8 \pi G  T^{matter}_{ab},
\label{EA}
\eq
where 
\bqn
G^{Einstein}_{ab} &=& R_{ab} - \frac{1}{2} g_{ab} R, \nb \\
T^{aether}_{ab}&=& \nabla_c [ J^c\;_{(a} u_{b)} + u^c J_{(ab)} - J_{(a} \;^c u_{b)}] - \frac{1}{2} g_{ab} J^c\;_d \nabla_c u^d+ \lambda u_a u_b  \nb \\
& & + c_1 [\nabla_a u_c \nabla_b u^c - \nabla^c u_a \nabla_c u_b] + c_4 a_a a_b, \nb \\
T^{matter}_{ab} &=&  \frac{- 2}{\sqrt{-g}} \frac{\delta \left( \sqrt{-g} L_{matter} \right)}{\delta g_{ab}}.
\label{fieldeqs}
\eqn

In a more general situation, the Lagrangian of GR theory is recovered, if and only if, the 
coupling constants are identically null, e.g., $c_1=c_2=c_3=c_4=0$, 
considering the equations
(\ref{Kab}) and (\ref{LagMul}).

\section{Spherical Solutions of EA field equations}

We start with the most general spherically symmetric static metric
\bq
ds^2= -e^{2A(r)} dt^2+e^{2B(r)} dr^2 +r^2 d\theta^2 +r^2 \sin^2 \theta d\phi^2.
\lb{ds2}
\eq
In accordance with equation (\ref{LagMul}), the aether field is assumed to be unitary and timelike, chosen as
\bq
u^a=(e^{-A(r)},0,0,0).
\eq
This choice is not the most general and is restricted to the scenario where aether is static. The aether must tip in a black hole solution as it cannot be timelike be aligned with the null Killing vector on the horizon. As that is not the case with our choice, our solutions are valid only outside the Killing horizon. This is good enough for solar system tests and even for astrophysical solutions  to describe the exterior spacetime to a source.

The timelike Killing vector of the metric (\ref{ds2}) is giving by
\bq
{\chi}^a = (-1, 0, 0, 0).
\eq
The Killing and the universal horizon \cite{Wang} \cite{Berglund2012} are obtained 
finding the largest root of
\bq
{\chi}^a {\chi}_a = 0,
\eq
and
\bq
{\chi}^a {u}_a = 0,
\eq
respectively, where ${\chi}^a$ is the timelike Killing vector. In our case, 
\bq
{\chi}^a {\chi}_a =-e^{2A(r)},
\lb{rkh}
\eq
\bq
{\chi}^a {u}_a = e^{A(r)}.
\lb{ruh}
\eq

For the general spherically symmetric metric (\ref{ds2}), we compute the different terms in the field equations (\ref{fieldeqs}) below.
\bqn
G^{aether}_{tt}=&&-\frac{e^{2(A-B)}}{2r^2}\left[ c_{14}(-2 r^2 A' B'+r^2 A'^2+2 r^2 A''+4 r A')\right.\nb\\
&&\left.~~~~~~~~~~~~~~~ -4 r B'-2 e^{2B}+2\right]=0,
\lb{Gtt}
\eqn

\bq
G^{aether}_{rr}=-\frac{1}{2r^2}\left( c_{14} r^2  A'^2 + 4 r A'- 2 e^{2B} + 2\right) = 0,
\lb{Grr}
\eq

\bq
G^{aether}_{\theta\theta}=\frac{r}{2e^{2B}}\left(2 r A'' - 2 r A' B'+ 2 A'- 2 B' -  (c_{14}-2) r A'^2 \right)=0,
\lb{Gthetatheta}
\eq
\bq
G^{aether}_{\phi\phi}=G^{aether}_{\theta\theta} \sin^2 \theta,
\eq
where $G^{aether}_{\mu\nu}=G^{Einstein}_{\mu\nu}-T^{aether}_{\mu\nu}$.
In order to identify eventual singularities in the solutions, it is useful to calculate the Kretschmann scalar invariant K. For the metric (\ref{ds2}), it is  given by
\bqn
K &=& \frac{4}{r^4 e^{4B}} \left(2 B'^2 r^2+e^{4B}-2 e^{2B}+1+2 A'^2 r^2+r^4 A''^2 + 2 r^4 A'' A'^2\right. \nb \\
&&\left.~~~~~~~~~ -2 r^4 A'' B' A'+r^4 A'^4-2 r^4 A'^3 B'+r^4 B'^2 A'^2\right).
\lb{K}
\eqn

Before we proceed, we simplify the field equations. Substituting the field equation (\ref{Grr}) into (\ref{Gtt}) we can eliminate the term $e^{2B}$ and find 
\bq
c_{14} \left(2rA''-2rA'B'+4A'\right)-4A'-4B'=0.
\lb{eq100}
\eq
We have two field equations, (\ref{Gthetatheta}) and (\ref{eq100}), to solve. Using these two, we can eliminate $A''$ only when $c_{14} \neq 0$ and  obtain the following,
\bq
(c_{14}-2) \left(B'+A'+\frac{1}{2} r c_{14}A'^2\right)=0.
\lb{eq1}
\eq
We can obtain another equation when $c_{14} = 0,$
\bq
r A''+ 2 r A'^2 + 2A'=0.
\eq
From the equation (\ref{eq1}) we can note that there are two possibilities: $c_{14} = 2$ and $c_{14} \neq 2$. As we mentioned previously, $ c_{14} > 2$  implies that the gravity is repulsive and $ c_{14} < 0$ means gravity is stronger. Thus, both mathematical and physical justifications lead us to four possible cases of interest: (i)  $ 0 < c_{14} < 2$ (ii) $c_{14} < 0$ (iii) $c_{14} = 2$ (iv) $c_{14} = 0$. Let us now analyze these four possible cases in detail.

In order to find explicit solutions for the metric functions we used
the Maple 16 algebraic software following the algorithm: we have made a survey of all
possible explicit solutions imposing $c_{14}=m/n$, where $m$ and $n$ are integer constants in the
interval $0<n<100$ and $0<m<100$ varying in step of one and imposing that $0<c_{14}<2$ or $c_{14}<0$. This survey has taken several hours of computational time in an old CPU generation
computer (Intel Duo Core).

\section{{\bf Solutions for} $\bf 0 < c_{14} < 2$ and $\bf c_{14} < 0$ }

The non-trivial solutions in EA theory exist for $ 0 < c_{14} < 2$ which help us understand the physical effects of aether. 
As we mentioned earlier, we have two field equations, (\ref{Gthetatheta}) and (\ref{eq100}), to solve. We first use (\ref{eq100}) to express $B'$ as
\bq
(2 + c_{14}r A' )B' =  c_{14} r A'' + 2 c_{14} A' -2 A'.
\eq
From this equation we can notice that if $2 + c_{14}r A'=0$, we would get  $A' = -\frac{2}{c_{14}r}$ which would not lead to any solution. Therefore, we proceed by assuming $2 + c_{14}r A' \neq 0$. 
Substituting  $B'$ into (\ref{Gthetatheta}) we have also
\bq
(c_{14}-2) \left(2 r A''+4 r A'^2+4 A'+r^2 c_{14}A'^3\right)=0.
\lb{eq2}
\eq
From the equations (\ref{eq1}) and (\ref{eq2}), we have
\bq
B'+A'+\frac{1}{2} r c_{14}A'^2=0,
\lb{eq1a}
\eq
and
\bq
2 r A''+4 r A'^2+4 A'+r^2 c_{14}A'^3=0. 
\lb{eq2a}
\eq

Substituting $B'$ and $A''$ from these equations into the field equations
(\ref{Gtt})-(\ref{Grr}) we get only an unique equation
\bq
r^2 c_{14} A'^2+4 r A'-2 e^{2B}+2=0.
\lb{Grr1}
\eq
Solving equation (\ref{Grr1}) for B we have
\bq
B=\frac{1}{2} \ln\left(\frac{1}{2} r^2A'^2 c_{14}+2A' r+1\right).
\lb{Bs}
\eq
Now, differentiating (\ref{Bs}) and substituting into (\ref{eq1a}) we get (\ref{eq2a}) which implies consistency.

\subsection{\bf Solution for ${\bf c_{14}=3/2}$}

Solving the equation (\ref{eq2a}) we can obtain an analytical solution for $A'$ for a particular case $c_{14}=3/2$.  
Making a transformation $u(r)=A'$ we can reduce the second order differential equation in $A$ into a first order differential equation in $u$, thus
\bq
2 r u'+4 r u^2+4 u+\frac{3}{2}r^2 u^3=0.
\lb{ur}
\eq
This equation is easily integrated giving three solutions which one is real and two are
imaginaries. The real one is given by
\bqn
u&=&A' = - \frac{2}{3r} \left\{ 1 -
\frac{\left[ r^2 \left( \sqrt{27 \beta^2-r^2}+3 \sqrt{3} \beta \right)\right]^{\frac{1}{3}}}
{\sqrt{27 \beta^2-r^2}}\right.\nb\\
&&\left. ~~~~~ ~~~~~  + \frac{r^2}{ \sqrt{27 \beta^2-r^2} \left[ r^2  \left(\sqrt{27 \beta^2-r^2}+3 \sqrt{3} \beta \right) \right]^{\frac{1}{3}}}\right\},
\lb{A1u}
\eqn
where $\beta$ is an arbitrary integration constant.

Since $r^2 < 27\beta^2$, we notice that this solution is not real in all spacetime but only in the region near to the center, meaning that this solution can not represent the exterior of a source.
Thus, it will not be analyzed in more detail.

\subsection{\bf Solution for ${\bf c_{14}=16/9}$}

Solving the equation (\ref{eq2a}) we can obtain an analytical solution for $A'$ for another particular case $c_{14}=16/9$.  Making a transformation $u(r)=A'$ we can reduce the second order differential equation in $A$ into a first order differential equation in $u$, thus
\bq
2 r u'+4 r u^2+4 u+\frac{16}{9}r^2 u^3=0.
\lb{ura}
\eq
This equation is easily integrated to get two solutions given by
\bqn
u&=&A' = - \frac{3}{4} \left(\frac{1}{r}+\frac{\delta} {\sqrt{(r-8 \alpha )r}}\right),
\lb{A1}
\eqn
where $\alpha$ is an arbitrary integration constant and $\delta=\pm 1$. Integrating once this equation and assuming the new arbitrary constant of integration to be zero,
without any loss of generality, we obtain
\bqn
A = -\frac{3}{4} \ln(r) -\frac{3}{4} \delta \ln\left[r-4 \alpha+\sqrt{(r-8 \alpha) r}\right].
\lb{Ar}
\eqn

When we calculate the limit $r \rightarrow +\infty$ we get
$e^{2A} \rightarrow 0$ when $\delta=+1$ and $e^{2A} \rightarrow 2 \sqrt{2}$ when $\delta=-1$. 
Thus, since we must have a flat spacetime at this limit, hereinafter, we assume $\delta =-1$.
Besides, we can also notice that for $\alpha>0$ the metric function (\ref{Ar}) becomes
imaginary for $r<8\alpha$. Thus, we assert that this solution could be 
used as the exterior vacuum solution of a static system, where $r>r_\Sigma=8\alpha$ and
$r_\Sigma$ would be the radius of the static system. 
In this way we can eliminate the pathological imaginary part
of the interior spacetime that has no physical meaning.
The case $\alpha=0$ gives us a flat spacetime since $A=\frac{3}{4}\ln(2)$ and $B=0$.
Thus, we will assume, hereinafter, $\alpha \le 0$.

We can now obtain $A''$ just differentiating equation (\ref{A1}). Thus,
\bqn
A'' = \frac{3}{4} \frac{\left[(r-4 \alpha)r+(8 \alpha -r)\sqrt{(r-8 \alpha) r}\right]}
{\left[r^2 (8 \alpha-r) \sqrt{(r-8 \alpha) r}\right]}.
\eqn
With the analytical solution $A'$ we can get $B$ from equation (\ref{Bs}) giving
\bqn
B &=& -\frac{1}{2} \ln(2)+\frac{1}{2} \ln\left(\sqrt{\frac{r}{r-8 \alpha}} + \frac{r}{r-8 \alpha} \right) .
\lb{Br}
\eqn

We can also obtain $A''$ and $B'$ analytically.
When we calculate the limit $r \rightarrow +\infty$ we get $e^{2B} \rightarrow 1$. 
We obtain $B'$ just differentiating equation (\ref{Br}), that is,
\bqn
B'=\frac{2 \alpha \left[r-8 \alpha+2 \sqrt{(r-8 \alpha) r}\right]}{r (8 \alpha-r) \left[r-8 \alpha+\sqrt{(r-8 \alpha) r}\right]}.
\eqn
In order to check if the analytical equations for 
$A'$, $B$, $A''$ and $B'$, we substitute them into the field equations (\ref{Gtt})-(\ref{Gthetatheta}) and we show that they are identically satisfied.
Thus, we can get analytically the metric components $g_{rr}=e^{2B}$,  $g_{tt}=e^{2A}$ 
and the Kretschmann scalar using equation (\ref{K}). The Kretschmann scalar obtained is 
\bq
K=\frac{768 \alpha^2\sqrt{r-8 \alpha}\left[2r(r-8\alpha)^2+(384 \alpha^2-104 r \alpha+7 r^2)\sqrt{(r-8 \alpha) r}\right]}{\left[r^{11/2}\left(r-8 \alpha+\sqrt{(r-8 \alpha) r}\right)^4\right] }.
\lb{Kra}
\eq
From the Kretschmann scalar, we can get the singularities which are at
\bqn
r_{sing1} = 0,\nb\\
r_{sing2} = 8 \alpha.
\eqn
Since the radial coordinate is always positive, the second singularity does not exist, since {$\alpha \le 0$}. 
So, the singularity at $r_{sing1}$ is physical and
\bqn
\lim_{r \rightarrow r_{sing1}^-} K = +\infty,\nb\\
\\
\lim_{r \rightarrow r_{sing1}^+} K = -\infty,
\eqn
which is independent of $\alpha$. 

Substituting this equation into (\ref{rkh}) and (\ref{ruh})
we get
\bq
{\chi}^a {\chi}_a =
-r^{-\frac{3}{2}}\left( r-4\alpha+\sqrt{(r- 8 \alpha) r}\right)^{\frac{3}{2}}=0,
\eq
\bq
{\chi}^a {u}_a =
r^{-\frac{3}{4}}\left( r-4\alpha+\sqrt{(r- 8 \alpha) r}\right)^{\frac{3}{4}}=0.
\eq
We can see easily again these equations do not have any root with $\alpha < 0$, 
hence, there exists no horizon, neither Killing nor universal horizon. 

In order to compare this result with the Schwarzschild spacetime, we will
assume here, for the sake of simplicity, $\alpha=-1$. From the Schwarzschild we have
\bqn
A_{Sch} = \frac{1}{2}\ln\left(1-\frac{2M}{r}\right),\nb\\
\\
B_{Sch} = -\frac{1}{2}\ln\left(1-\frac{2M}{r}\right),
\lb{Sch}
\eqn
where $M$ is the mass of the particle. Substituting these metric functions into the equation
(\ref{K}) we have the well known Kretschmann scalar for the Schwarzschild metric
\bq
K_{Sch}=\frac{48M^2}{r^6}.
\lb{Ksch}
\eq
 
From equations (\ref{Ar}) and (\ref{Br}) we get $A$ and $B$ for $\alpha=-1$. From this equation we can get $g_{rr}=e^{2B}$ 
and $g_{tt}=e^{2B}$ analytically.
\bqn
A&=&-\frac{3}{4} \ln(r) + \frac{3}{4} \ln{\left[r+4+\sqrt{(r+8) r}\right]},\nb\\
\\
B &=& -\frac{1}{2} \ln(2)+\frac{1}{2} \ln\left(\sqrt{\frac{r}{r+8}} + \frac{r}{r+8} \right) .
\eqn
The Kretschmann scalar is obtained from equation (\ref{Kra}) $K$ for $\alpha=-1$, thus
\bq
K=\frac{768 \sqrt{r+8}\left[2r(r+8)^2+(384+104 r +7 r^2)\sqrt{(r+8) r}\right]}{\left[r^{11/2}\left(r+8 +\sqrt{(r+8) r}\right)^4\right] }.
\eq

\begin{figure}[!ht]
\centering	
\begin{minipage}[t]{1\textwidth}
\centering	
\includegraphics[width=0.6\linewidth]{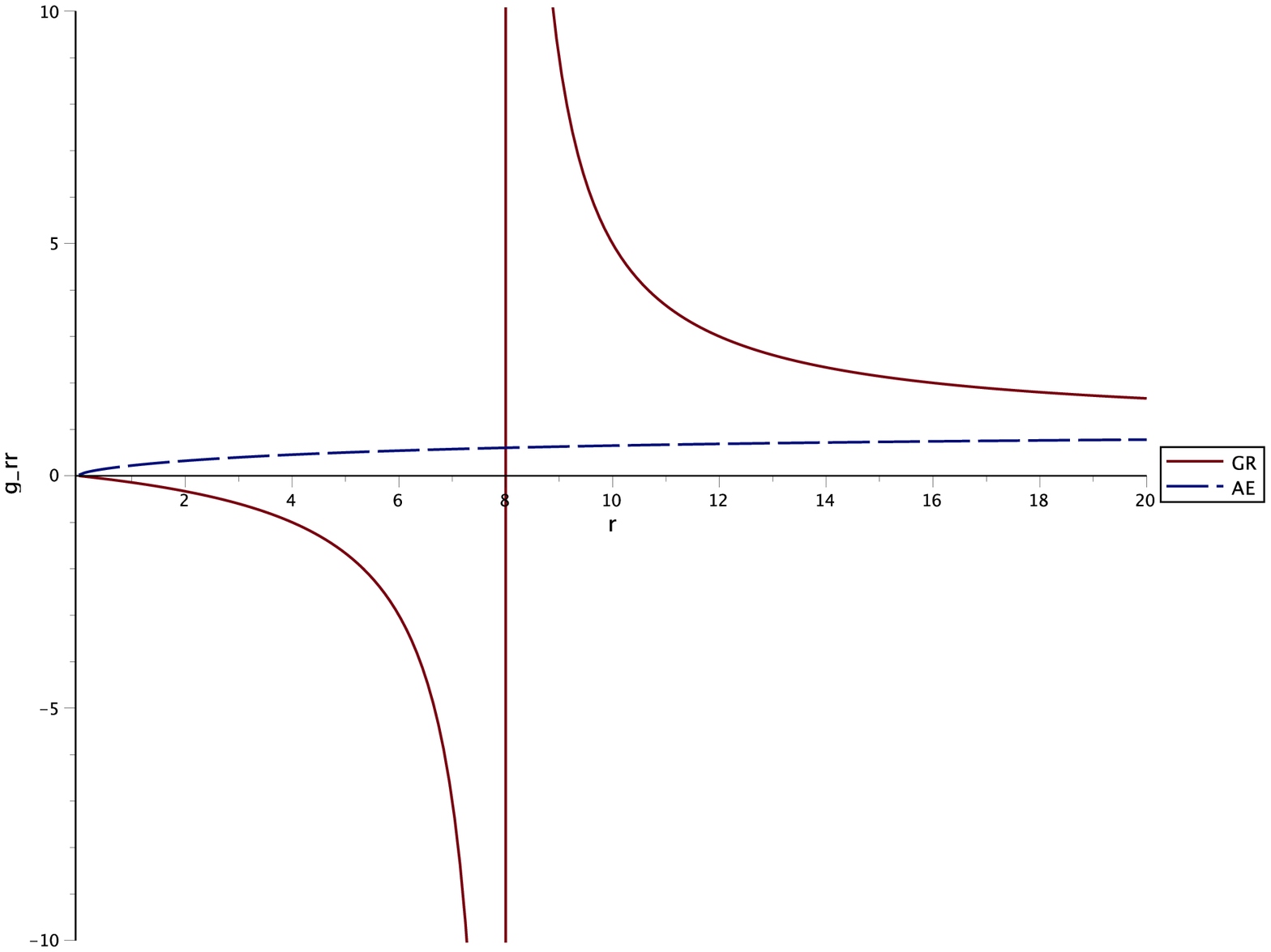}
\caption{Plot of the metric component $g_{rr}$, for EA parameters 
$c_{14}=16/9$, $\alpha=-1$
and an arbitrary Schwarzschild mass $M=4$. 
The continuous red line
represents the GR Schwarzschild metric. The dashed blue line represents the EA solution. 
Calculating the limit $r \rightarrow +\infty$ we obtain $g_{rr}=e^{2B} \rightarrow 1$, as we can see in this figure. 
Calculating the limit $r \rightarrow r_{sing1}=0$ we obtain $g_{rr}=e^{2B} \rightarrow 0$, as we can see in this figure.} 
\label{fig5}
\end{minipage}
\end{figure}

\begin{figure}[!ht]
	\centering	
\begin{minipage}[t]{1\textwidth}
\centering	
\includegraphics[width=0.6\linewidth]{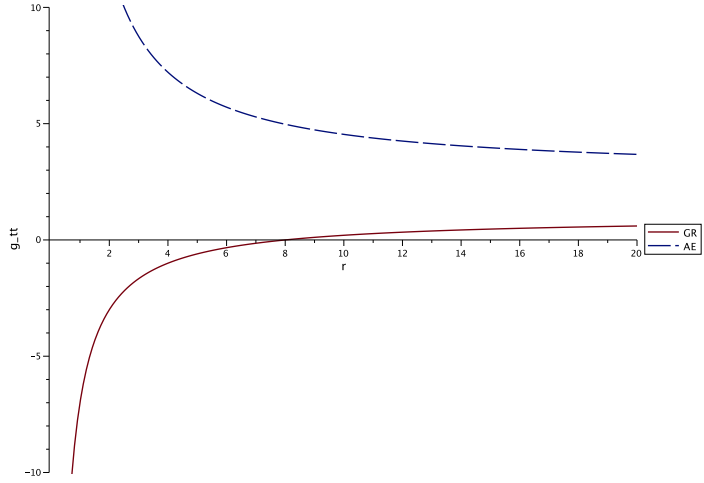}
\caption{Plot of the metric component $g_{tt}$, for EA parameters $c_{14}=16/9$, $\alpha=-1$  and an arbitrary Schwarzschild mass $M=4$. 
The continuous red line
represents the GR Schwarzschild metric. The dashed blue line represents the EA solution. Calculating the limit $r \rightarrow +\infty$ we obtain $g_{tt}=e^{2A} \rightarrow 2\sqrt{2}$, as we can see in this figure.}
\label{fig6}
\end{minipage}
\end{figure}

\begin{figure}[!ht]
	\centering	
\begin{minipage}[t]{1\textwidth}
\centering	
\includegraphics[width=0.6\linewidth]{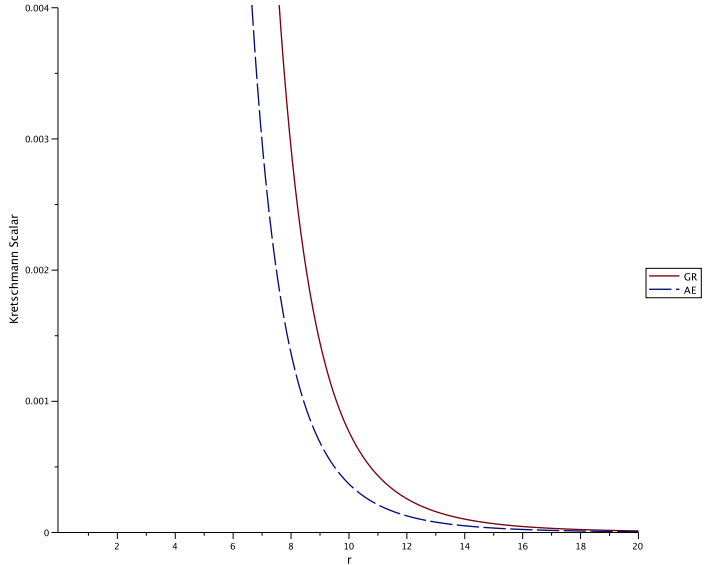}
\caption{Plot of the Kretschmann scalar, for EA parameters $c_{14}=16/9$, 
$\alpha=-1$ 
and an arbitrary Schwarzschild mass $M=4$. 
The continuous red line
represents the GR Schwarzschild metric. The dashed blue line represents the EA solution. Calculating the limit $r \rightarrow +\infty$ we obtain 
$K \rightarrow 0$. 
Calculating the limit from left $r \rightarrow r_{sing1}=0$ we get
$K \rightarrow  -\infty$, as we can see in this figure. Calculating the limit from right $r \rightarrow r_{sing1}=0$ we get
$K \rightarrow  +\infty$, as we can see in this figure.}
\label{fig7}
\end{minipage}
\end{figure}

In the Figures \ref{fig5}-\ref{fig7} we show the comparison of the present work
quantities $g_{rr}$, $g_{tt}$ and $K$ with the Schwarzschild metric ones.

\subsection{\bf Solution for ${\bf c_{14}=48/25}$}

Solving the equation (\ref{eq2a}) we can obtain an analytical solution
for $A'$ for another particular case $c_{14}=48/25$. Making a transformation $u(r)=A'$
we can reduce the second order differential equation in $A$ into a first order 
differential equation in $u$, thus
\bq
2 r u'+4 r u^2+4 u+\frac{48}{25}r^2 u^3=0.
\lb{uraa}
\eq
This equation is easily integrated giving 
\bqn
&& u = A' = -\frac{5\;.\;2^{\frac{1}{3}}}{12r(-2 r+27 \gamma)} \left[ r \left(1-3 \sqrt{3}\sqrt{\frac{\gamma}{-2 r+27 \gamma}} \right) (-2 r+27 \gamma)^2 \right]^{\frac{1}{3}}\nb\\
&& ~~~~~~~~~~ +{\frac{5\;.\; 2^{\frac{2}{3}}}{12}} \left[r \left( 1-3 \sqrt{3}\sqrt{\frac{\gamma}{-2 r+27 \gamma}} \right) (-2 r+27 \gamma)^2 \right]^{-\frac{1}{3}}-{\frac{5}{6r}},
\lb{A1a}
\eqn
where $\gamma$ is an arbitrary integration constant.
Substituting equation (\ref{A1a}) into (\ref{Bs}) we can calculate the limit 
$r \rightarrow +\infty$ we get $e^{2B} \rightarrow 1$ only when $\gamma<0$.
Thus, since we must have a flat spacetime at this limit, hereinafter, we assume 
$\gamma =-|\gamma|$. Thus,
\bqn
u&=&A' = \frac{5}{6} \left\{ - r^{-1} +  (2 r)^{-\frac{2}{3}} (2 r+27 |\gamma|)^{-\frac{1}{2}} \left(\sqrt{2 r+27|\gamma|} - \sqrt{27|\gamma|} \right)^{\frac{1}{3}}  \right.\nb\\
&& ~~~~~~~~~~~ \left. +  (2 r)^{-\frac{1}{3}} (2 r+27 |\gamma|)^{-\frac{1}{2}} \left(\sqrt{2 r+27|\gamma|} - \sqrt{27|\gamma|} \right)^{-\frac{1}{3}}   \right\}.
\lb{A1b}
\eqn

Substituting equation (\ref{A1b}) into (\ref{Bs}) we get
	\bqn
	&&B = -{\frac{1}{3}} \ln(2)-{\frac{1}{2}} \ln(3)+{\frac{1}{2}} \ln \left[ 2^{\frac{5}{3}} r \left(2r+27|\gamma| \right)^{-1} \right.\nb\\
	&& ~~~~~~
	+ 2 r^{\frac{4}{3}} \left(2r + 27|\gamma|\right)^{-1} \left( \sqrt{27|\gamma|}-\sqrt{2 r+27|\gamma|} \right)^{-\frac{2}{3}} 	\nb\\
	&& ~~~~~~
	- 2^{\frac{1}{3}} r^{\frac{2}{3}} (2 r+27 |\gamma|)^{-\frac{1}{2}} \left( \sqrt{27|\gamma|}-\sqrt{2 r+27|\gamma|} \right)^{-\frac{1}{3}}  
	\nb\\
	&& ~~~~~~
	+ 2^{\frac{1}{3}} r^{\frac{2}{3}} (2 r+27|\gamma|)^{-1} \left( \sqrt{27|\gamma|}-\sqrt{2 r+27|\gamma|} \right)^{\frac{2}{3}}  
	\nb\\
	&& ~~~~~~ \left.	
	- r^{\frac{1}{3}} (2 r+27 |\gamma|)^{-\frac{1}{2}} \left( \sqrt{27|\gamma|}-\sqrt{2 r+27|\gamma|} \right)^{\frac{1}{3}} \right].
	\eqn

Integrating the equation (\ref{A1b}) using \textit{Mathematica} for $|\gamma| = 1$ , we obtain $A$ which is given by
\bqn
A&=&  \frac{5}{4 (2 r)^{2/3}}  \left(\sqrt{2 r+27}- \sqrt{27}\right)^{4/3} \,
_2F_1\left(\frac{2}{3},1;\frac{5}{3};\frac{r-3 \sqrt{6 r+81}+27}{r}\right) 
\nb \\ && 
- \frac{5}{6} \log (r) - \frac{5}{6 (4 r)^{2/3}}  \left(3 \sqrt{6r+81}+27+r\right)^{1/3} \left(\sqrt{2 r+27}-\sqrt{27}\right)^{2/3} \nb \\
&& 
\times \left[ \log
\left(\sqrt{2 r+27}- \sqrt{27}\right)+3 \log \left(1-\sqrt[3]{\frac{3 \sqrt{6 r+81}+27}{r}+1}\right) \right. \nb \\
&& ~~~~~ \left.	+2 \sqrt{3} \tan ^{-1}\left(\frac{2 \sqrt[3]{\frac{3 \sqrt{6 r+81}+27}{r}+1}+1}{\sqrt{3}}\right)\right] 
\eqn
where $_2F_1(a,b;c;x)$ is a hypergeometric function. Substituting this equation into (\ref{rkh}) and (\ref{ruh}) we found that these equations do not have any root, hence, there exists no horizon, neither Killing nor universal. Thus, we have naked singularities.

We obtain $B'$ just differentiating equation (\ref{Br}). In order to check the consistency of 
$A'$, $B$, $A''$ and $B'$, we substitute them into the field equations (\ref{Gtt})-(\ref{Gthetatheta}) and show that they are identically satisfied. Now, we can get analytically the metric components $g_{rr}=e^{2B}$,  $g_{tt}=e^{2A}$ 
and the Kretschmann scalar using equation (\ref{K}). 
The Kretschmann scalar is calculated but not shown because it is too long.
From the Kretschmann scalar we can get the singularity which are given by
\bqn
r_{sing1} = 0.
\eqn

\begin{figure}[!ht]
\centering	
\begin{minipage}[t]{1\textwidth}
\centering	
\includegraphics[width=0.6\linewidth]{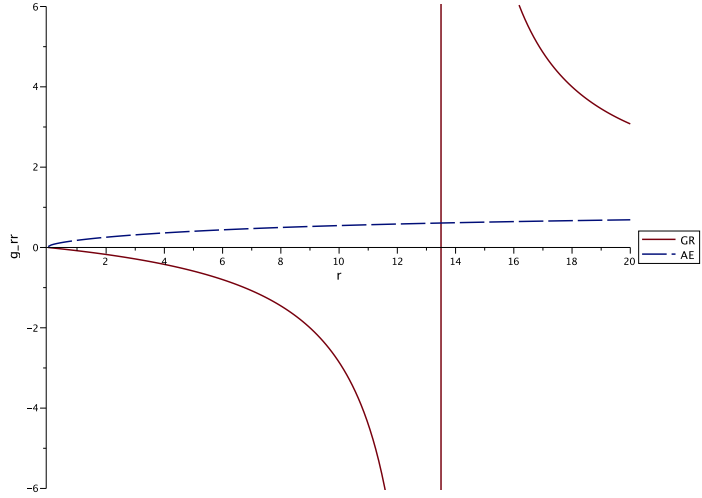}
\caption{Plot of the metric component $g_{rr}$, for EA parameters 
$c_{14}=48/25$, $|\gamma|=1$
and an arbitrary Schwarzschild mass $M=27/4$. The continuous red line
represents the GR Schwarzschild metric. The dashed blue line represents the EA solution.
Calculating the limit $r \rightarrow +\infty$ we obtain $g_{rr}=e^{2B} \rightarrow 1$, as we can see in this figure. 
Calculating the limit $r \rightarrow r_{sing1}=0$  we obtain $g_{rr}=e^{2B} \rightarrow 0$, as we can see in this figure.}
\label{fig8}
\end{minipage}
\end{figure}

\begin{figure}[!ht]
	\centering	
\begin{minipage}[t]{1\textwidth}
\centering	
\includegraphics[width=0.6\linewidth]{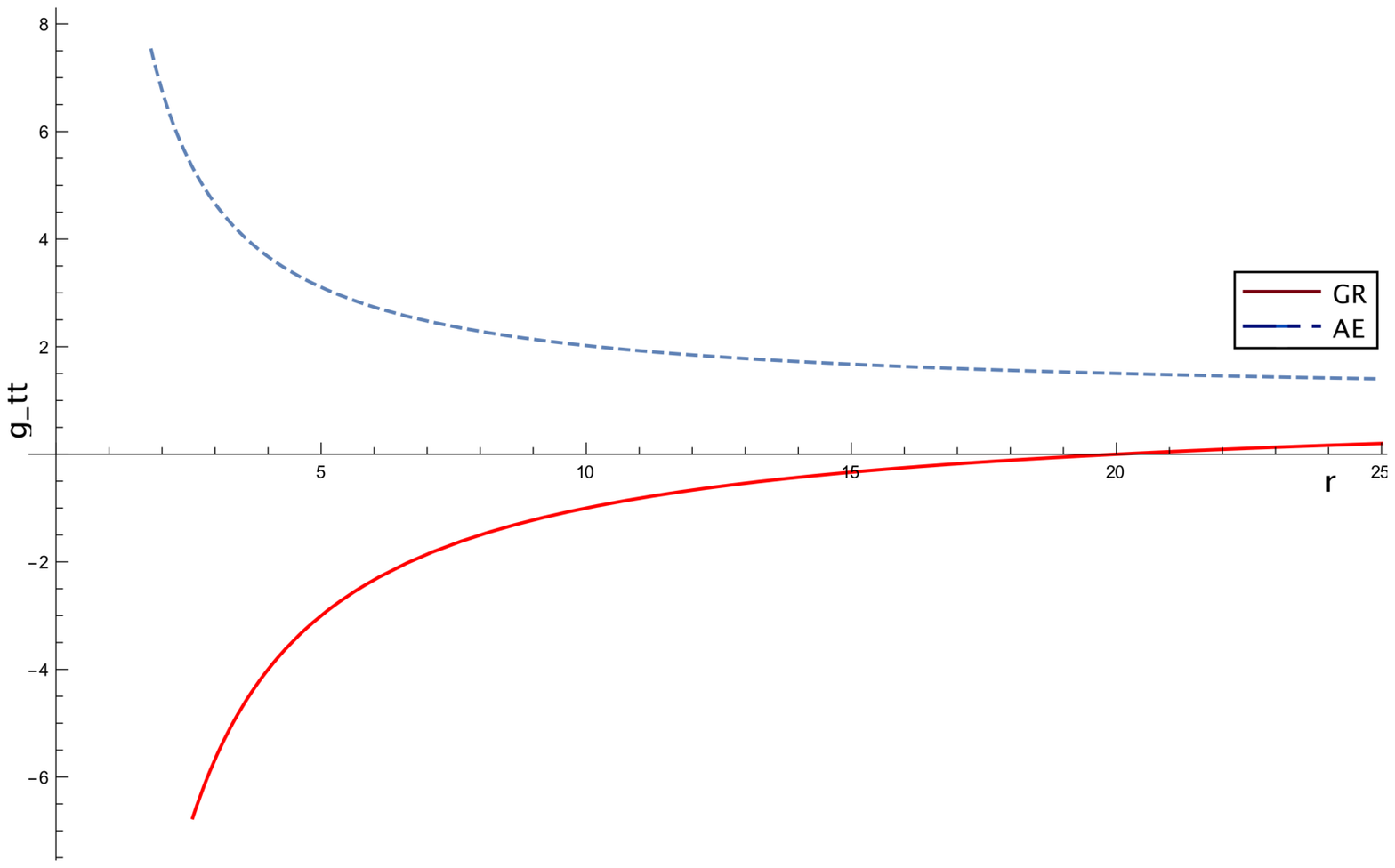}
\caption{Plot of the metric component $g_{tt}$, for EA parameters 
$c_{14}=48/25$, $|\gamma|=1$ and an arbitrary Schwarzschild mass $M=10$
The continuous red line
represents the GR Schwarzschild metric. The dashed blue line represents the EA solution. Calculating the limit $r \rightarrow +\infty$,
we obtain $g_{tt}=e^{2A} \rightarrow 1$, as we can see in this figure. 
Calculating the limit from left $r \rightarrow r_{sing1}=0$ 
we obtain $g_{tt}=e^{2A} \rightarrow +\infty$, as we can see in this figure.}
\label{fig9}
\end{minipage}
\end{figure}

\begin{figure}[!ht]
	\centering	
\begin{minipage}[t]{1\textwidth}
\centering	
\includegraphics[width=0.6\linewidth]{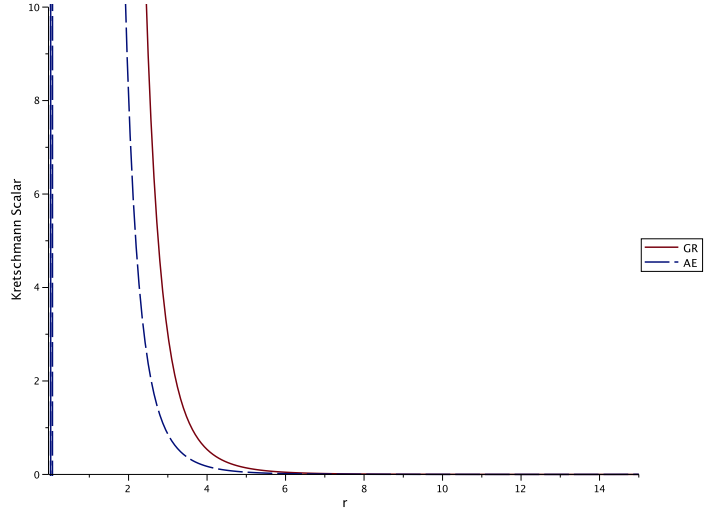}
\caption{Plot of the Kretschmann scalar, for EA parameters $c_{14}=48/25$, 
$|\gamma|=1$ and an arbitrary Schwarzschild mass $M=27/2$. 
The continuous red line
represents the GR Schwarzschild metric. The dashed blue line represents the EA solution. 
Calculating the limit $r \rightarrow +\infty$ we obtain $K \rightarrow 0$. 
Calculating the limit from left $r \rightarrow r_{sing1}=0$ we obtain $K \rightarrow  -\infty$, as we can see in this figure. 
Calculating the limit from right $r \rightarrow r_{sing1}=0$ we obtain $K \rightarrow  +\infty$, as we can see in this figure.}
\label{fig11}
\end{minipage}
\end{figure}

In order to compare this result with the Schwarzschild spacetime, we use 
the equations (\ref{Sch}) and (\ref{Ksch}). For the sake of simplicity we assume also 
$|\gamma|=1$. 
The Kretschmann scalar is obtained from equation (\ref{K}) analytically. 
In the Figures \ref{fig8}-\ref{fig11} we show the comparison of the present work
quantities $g_{rr}$, $g_{tt}$ and $K$ with the Schwarzschild metric ones.

\subsection{\bf Solution for ${\bf c_{14} < 0}$}

{
Taking $c_{14}<0$ implies that the coupling constant of EA theory is stronger than the Newtonian gravitational constant [see equation (\ref{Ge})]}. Although this possibility is ruled out by experimental data \cite{Oost2018}, we include this case just for the sake of completeness. Solving the equation (\ref{eq2a}) we can obtain an analytical solution
for $A'$ for another particular case $c_{14}=-16$. Making a transformation $u(r)=A'$
we can reduce the second order differential equation in $A$ into a first order 
differential equation in $u$, thus
\bq
2 r u'+4 r u^2+4 u-16r^2 u^3=0.
\lb{urab}
\eq
This equation is easily integrated giving

\bqn
u=A'= \frac{ \kappa\left[ \left(4+4r \sqrt{\frac{r}{-32 \kappa^3+r^3}}\right)^{\frac{2}{3}} \left(32 \kappa^3-r^3\right)^{\frac{1}{3}} +8 \kappa\right]}{2  r \left(32 \kappa^3-r^3\right)^{\frac{2}{3}} \left( 4+4r \sqrt{\frac{r}{-32 \kappa^3+r^3}} \right)^{\frac{1}{3}}},\nb\\
\lb{A1-16a}
\eqn
where $\kappa$ is an arbitrary integration constant. The solution with $\kappa>0$ give
us an imaginary solution, thus it will not be considered here. The solution with $\kappa<0$
is gives us real solution. Thus in the following equation we assume that $\kappa=-|\kappa|$.

\bqn
u=A'&&= \frac{ 2^{\frac{5}{9}} |\kappa| \left( \sqrt{32 |\kappa|^3+r^3} + \sqrt{r^3}\right)^{\frac{1}{3}} }{2  r \sqrt{32 |\kappa|^3+r^3}  } \nb \\
&& 
+ \frac{ 2 |\kappa|}{  r \sqrt{32 |\kappa|^3+r^3} \left( \sqrt{32 |\kappa|^3+r^3} + \sqrt{r^3}\right)^{\frac{1}{3}} }
\lb{A1-16}
\eqn

Substituting equation (\ref{A1-16}) into (\ref{Bs}) we can calculate the limit 
$r \rightarrow +\infty$ we get $e^{2B} \rightarrow 1$ only when $\kappa<0$.
Thus, we have a flat spacetime at this limit. 
Thus,
\bqn
B &=& {\frac{1}{3}} \ln(2)+{\frac{1}{2}} \ln \left\{ r \left\{ 2048 \sqrt{\frac{r}{32 |\kappa|^3+r^3}}) |\kappa|^7-\right. \right.\nb\\
&&\left. \left. 128\;.\; 2^{\frac{2}{3}}\;|\kappa|^5 \left[ \left( 1+r \sqrt{\frac{r}{32 |\kappa|^3+r^3}} \right)  (32 |\kappa|^3+r^3)^2 \right]^{\frac{1}{3}} \sqrt{\frac{r}{32 |\kappa|^3+r^3}} +\right. \right.\nb\\
&&\left. \left. 128\; |\kappa|^4 r^3 \sqrt{\frac{r}{32 |\kappa|^3+r^3}})+64\; |\kappa|^4 r^2-\right. \right.\nb\\
&&\left. \left. 4\;.\; 2^{\frac{2}{3}} |\kappa|^2\; r^3 \left[ \left( 1+r \sqrt{\frac{r}{32 |\kappa|^3+r^3}} \right)  (32 |\kappa|^3+r^3)^2 \right]^{\frac{1}{3}} \sqrt{\frac{r}{32 |\kappa|^3+r^3}} +\right. \right.\nb\\
&&\left. \left. 2 |\kappa|\; r^6 \sqrt{\frac{r}{32 |\kappa|^3+r^3}})+2 |\kappa| r^5+\right. \right.\nb\\
&&\left. \left. 2^{\frac{1}{3}}r^2 \left[ \left( 1+r \sqrt{\frac{r}{32 |\kappa|^3+r^3}} \right) 
(32 |\kappa|^3+r^3)^2 \right]^{\frac{2}{3}} \right\} \times \right. \nb\\
&&\left. \left[ \left( 4 |\kappa|+2^{\frac{1}{3}} r \right) \left( 16 |\kappa|^2-4 2^{\frac{1}{3}} |\kappa| r+2^{\frac{2}{3}} r^2 \right) \times \right. \right.\nb\\
&&\left. \left.  \left[ \left( 1+r \sqrt{\frac{r}{32 |\kappa|^3+r^3}} \right) (32 |\kappa|^3+r^3)^2 \right]^{\frac{2}{3}} \right]^{-1} \right\}.
\eqn

\begin{figure}[!ht]
\centering	
\begin{minipage}[t]{1\textwidth}
\centering	
\includegraphics[width=0.6\linewidth]{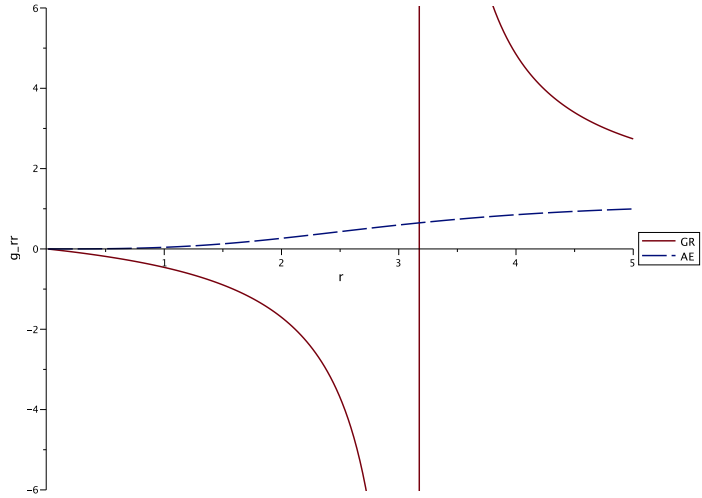}
\caption{Plot of the metric component $g_{rr}$, for EA parameters 
$c_{14}=-16$, $|\kappa|=1$
and an arbitrary Schwarzschild mass $M=32^{\frac{1}{3}}/2$. 
The continuous red line
represents the GR Schwarzschild metric. The dashed blue line represents the EA solution. 
Calculating the limit $r \rightarrow +\infty$ we obtain $g_{rr}=e^{2B} \rightarrow 1$, as we can see in this figure. 
Calculating the limit $r \rightarrow r_{sing1}=0$  we obtain $g_{rr}=e^{2B} \rightarrow 0$, as we can see in this figure.}
\label{fig12}
\end{minipage}
\end{figure}

\begin{figure}[!ht]
	\centering	
\begin{minipage}[t]{1\textwidth}
\centering	
\includegraphics[width=0.6\linewidth]{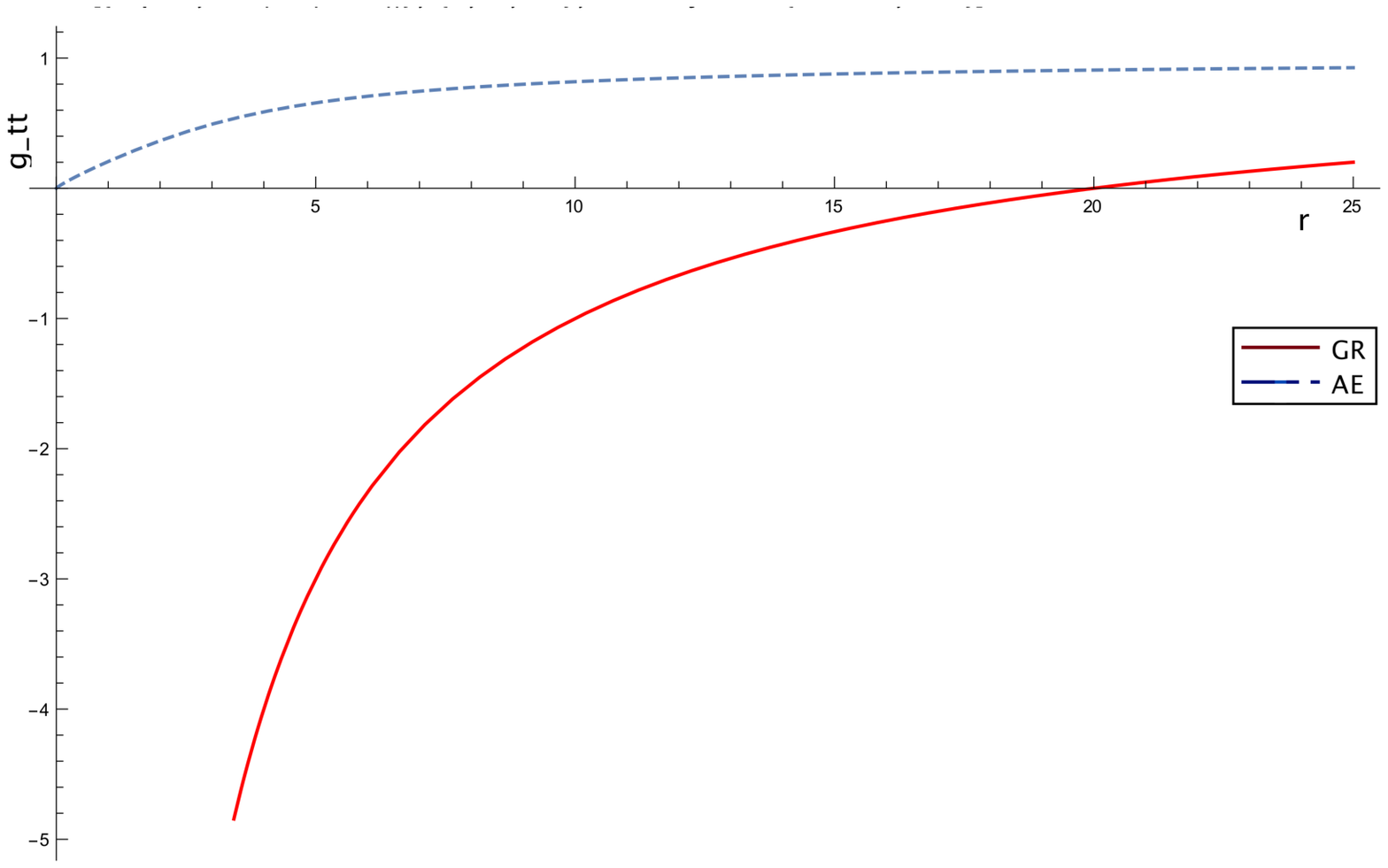}
\caption{	Plot of the metric component $g_{tt}$, for EA parameters 
$c_{14}=-16$, $|\kappa|=1$ and an arbitrary Schwarzschild mass $M=10$.
The continuous red line
represents the GR Schwarzschild metric. The dashed blue line represents the EA solution.  
Calculating the limit $r \rightarrow +\infty$, we obtain $g_{tt}=e^{2A} \rightarrow 1$, as we can see in this figure. 
Calculating the limit from {right} $r \rightarrow r_{sing1}=0$ 
we obtain $g_{tt}=e^{2A} \rightarrow {0}$, as we can see in this figure.}
\label{fig13}
\end{minipage}
\end{figure}

\begin{figure}[!ht]
	\centering	
\begin{minipage}[t]{1\textwidth}
\centering	
\includegraphics[width=0.6\linewidth]{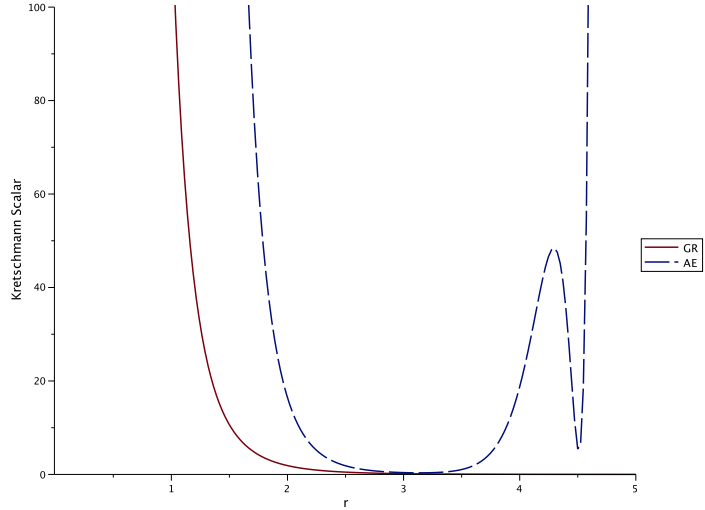}
\caption{Plot of the Kretschmann scalar, for EA parameters $c_{14}=-16$, 
$|\kappa|=1$ and an arbitrary Schwarzschild mass $M=32^{\frac{1}{3}}/2$.
The continuous red line
represents the GR Schwarzschild metric. The dashed blue line represents the EA solution.  
Calculating the limit $r \rightarrow +\infty$ we obtain $K \rightarrow +\infty$, as we can see in this figure. 
Calculating the limit from left $r \rightarrow r_{sing1}=0$ we obtain $K \rightarrow  +\infty$, as we can see in this figure.}
\label{fig15}
\end{minipage}
\end{figure}

Integrating Eq. (\ref{A1-16}) using \textit{Mathematica} for $|\kappa|=1$, we obtain
\bqn
A &=& \frac{1}{96\ 2^{4/9}} \left[ -3 \left(r^{3/2}+\sqrt{r^3+32}\right)^{4/3} \,
_2F_1\left(\frac{2}{3},1;\frac{5}{3};\frac{1}{16} \left(r^3+\sqrt{r^3+32}
r^{3/2}+16\right)\right) \right. \nb \\ 
&&	-8 \sqrt[9]{2} \left(-2 \log \left(2\
2^{2/3}-\left(r^{3/2}+\sqrt{r^3+32}\right)^{2/3}\right) \right. \nb \\ 
&&
+\log
\left(\left(r^{3/2}+\sqrt{r^3+32}\right)^{4/3} 
+2\ 2^{2/3} 	\left(r^{3/2}+\sqrt{r^3+32}\right)^{2/3}+8 \sqrt[3]{2}\right)  \nb \\
&&\left. \left.	
+2 \sqrt{3} \tan
^{-1}\left(\frac{\left(r^{3/2}+\sqrt{r^3+32}\right)^{2/3}+2^{2/3}}{2^{2/3}
	\sqrt{3}}\right)\right) \right],
\eqn
where $_2F_1(a,b;c;x)$ is a hypergeometric function. Substituting this equation into (\ref{rkh}) and (\ref{ruh}) we found that these equations do not have any root, hence, there exists no horizon, neither Killing nor universal. Thus, we have naked singularities.
In the Figures \ref{fig12}-\ref{fig15} we show the comparison of the present work
quantities $g_{rr}$, $g_{tt}$ and $K$ with the Schwarzschild metric ones.

\section{{\bf Solutions for $\bf c_{14} = 2$}}
%
Taking $c_{14}=2$ implies that the coupling constant of EA theory vanishes. This matterless case may not be of any physical interest, nevertheless we analyze this for the sake of completeness. 
Substituting this condition into field equations (\ref{Gtt})-(\ref{Gthetatheta}), we
get
\bq
2 r^2 A'' - 2 r^2 A' B' + r^2 A'^2 + 4 r A' - 2 r B' - e^{2B} + 1=0,
\lb{Gtt2}
\eq
\bq
r^2 A'^2 + 2 r A' - e^{2B} + 1 =0,
\lb{Grr2}
\eq 
\bq
r A'' - r  A' B' + A' - B'= 0.
\lb{Gthetatheta2}
\eq 
Equations (\ref{Grr2}) and (\ref{Gthetatheta2}) can be combined to get (\ref{Gtt2}). In fact, we can directly read off the solution  (\ref{Grr2}) which gives,
\bq
B =\ln\left[\pm(1+ r A')\right],
\lb{Bs1}
\eq
where $A$ is an arbitrary function of the $r$. Below, we show an example where we consider the metric function $g_{tt}$ to be the same as the one in Schwarzschild solution. That is,
\bq
A = \frac{1}{2}\ln\left(1-\frac{2M}{r}\right),
\eq
obtaining from (\ref{Bs1})
\bq
B = \frac{1}{2}\ln\left[\frac{(M-r)^2}{(2M-r)^2}\right],
\eq
where $M$ is a constant.

Substituting these equations into (\ref{rkh}) and (\ref{ruh}) we get
\bq
{\chi}^a {\chi}_a =-1+\frac{2M}{r}=0,
\eq
thus, the Killing horizon is at $r_{KH}=2M,$ and
\bq
{\chi}^a {u}_a =
\sqrt{1-\frac{2M}{r}}=0,
\eq
so, the universal horizon is also at $r_{UH}=2M.$ The Killing and the universal horizons coincide as in GR.

The metric for this case is given by
\bq
ds^2=-\left(1-\frac{2M}{r}\right)dt^2+\frac{(M-r)^2}{(2M-r)^2} dr^2 +r^2 d\theta^2 +r^2 \sin^2 \theta d\phi^2,
\eq
while the Kretschmann scalar is given by
\bq
K=\frac{4 M^2}{r^4 (M-r)^6} \left(26 M^4-90 r M^3+119 M^2 r^2-64 r^3 M+12 r^4\right),
\eq
and their limits are
\bq
\lim_{r \to M} K=+\infty,
\eq
\bq
\lim_{r \to 0} K=+\infty.
\eq
This spacetime is Minkowski at the infinity
\bq
\lim_{r \to +\infty} K = 0.
\eq

\section{{\bf Solutions for $\bf c_{14} = 0$}}
%
Taking $c_{14}=0$ implies that the coupling constant of EA theory is the same as the Newtonian gravitational constant G. Substituting $c_{14}=0$ into the field equations (\ref{Gtt})-(\ref{Gthetatheta}), we get
\bq
2 r B'+ e^{2B}-1 =0,
\eq
\bq
2 r A' -  e^{2B} +1 =0,
\eq
\bq
r A'' - r  A' B' + r A'^2 + A' - B'  =0.
\eq
From the above, we have the solution
\bq
A=\frac{1}{2}\ln{\left(2 - \frac{2\nu}{r}\right)},
\eq 
and 
\bq
B= -\frac{1}{2}\ln{{\left(1 -\frac{\nu}{\mu r}\right)}}.
\eq
where $\mu$ and $\nu$ are some constants of integration which can be chosen appropriately. With this solution for $A$ and $B$, we can cast the metric (\ref{ds2})  in the Schwarzschild form, that is,
\bq
ds^2= -2\left(\mu -\frac{\nu}{r}\right) dt^2+ \mu \left(\mu -\frac{\nu}{r}\right)^{-1} dr^2 +r^2 d\theta^2 +r^2 \sin^2 \theta d\phi^2,
\eq
which yields the standard Schwarzschild metric of GR when $\mu=1/2$ and $\nu=M$.

Substituting $A$ and $B$ equations into (\ref{rkh}) and (\ref{ruh}) we get
\bq
{\chi}^a {\chi}_a =-1+\frac{2M}{r}=0,
\eq
which means the Killing horizon is at $r_{KH}=2M,$ and
\bq
{\chi}^a {u}_a =
\sqrt{1-\frac{2M}{r}}=0,
\eq
implies the universal horizon is at $r_{UH}=2M.$ We can see again that the Killing and the universal horizons coincide just as in GR.

\section{Conclusions}

In the present work, we have analyzed all the possible exterior vacuum solutions with spherical symmetry allowed by the EA theory with static aether. We show that there are four classes of explicit analytic solutions corresponding to different values for $c_{14}$, which are: $0 <c_{14}<2$, $c_{14}<0$,  $c_{14}=2$ and $c_{14}=0$. Our results are summarized below.
\begin{enumerate}
	
	\item For $0 < c_{14} <2$ we present analytical solutions for $c_{14}=3/2$, $c_{14}=16/9$ and $c_{14}=48/25$.  The $c_{14}=3/2$ solution is excluded because the spacetime becomes imaginary at a finite radius. The other two analytical solutions ($c_{14}=16/9 \textrm{ and } 48/25$) have singularities at $r=0$, and are asymptotically flat spacetimes as in the case of Schwarzschild metric in GR. These solutions have no horizons, neither Killing nor universal. \textcolor{black}{This suggests that} we \textcolor{black}{can} have naked singularities. 

	\item For $c_{14} < 0$, \textcolor{black}{the absence of any horizons again suggests the presence of} a naked singularity at the origin and the spacetime is asymptotically flat. 
	
	\item For $c_{14}=2$ we have a family of new static solutions with an arbitrary function of the coordinate $r$. As a consequence of $c_{14}=2$ the EA coupling constant is zero, means that there is no coupling with the matter, which does not pose any problem to vacuum solutions. However, the difficulty  arises when matching the exterior vacuum  with some interior source, since the EA parameter should be the same in the whole spacetime (interior and exterior). This is hitherto unnoticed in the literature. When A(r) is chosen to be the same as in the Schwarzschild case, the solution for $c_{14}=2$ has two physical singularities, one at a finite radius $r=M$ besides the usual at the origin $r=0$. This characteristic is completely different from the Schwarzschild spacetime in GR. Both Killing and universal horizons in this case coincide with event horizon of the Schwarzschild spacetime in GR, i.e., $r=2M$.
		
	\item For $c_{14}=0$ we have the Schwarzschild spacetime, which also picks Newtonian gravitational constant as coupling. As expected, the Killing horizon coincides with the universal horizon and is the same as the event horizon of the Schwarzschild
	spacetime in GR.	
\end{enumerate}

In the 2006 paper, Eling and Jacobson \cite{Eling2006} analyzed the behavior of the static aether solutions for $0 \le c_{14} <2$. They asserted that regular black holes cannot have static aether fields since the Killing vector is null, not timelike on the horizon as always is the case with static aether.
\textcolor{black}{As far as we know, this the only work in EA theory that
mentions the existence of naked singularity.}
 What we showed here is the corollary, meaning the static aether leads to \textcolor{black}{absence of any kind of horizon, suggesting a possible formation of} naked singularities\textcolor{black}{, which could circumvent} the need for switching from timelike vector to null vector at the horizon by totally avoiding the existence of the horizon itself.  \textcolor{black}{Of course, to rigorously prove that the singularity is naked,  it would be still necessary to prove that  future directed null geodesics can emanate from the singularity.}

\section {Acknowledgments}

The author (RC) acknowledges the financial support from FAPERJ (no.E-26/171.754/2000, E-26/171.533/2002 and E-26/170.951/2006). MFAdaS  acknowledges the financial support from CNPq-Brazil, FINEP-Brazil (Ref. 2399/03), FAPERJ/UERJ (307935/2018-3) and from CAPES (CAPES-PRINT 41/2017). VHS thanks Jos\'{e} Abdalla Helay\"{a}l-Neto  for hospitality at Centro Brasileiro de Pesquisas F\'{i}sicas (CBPF). The authors are grateful to Ted Jacobson for pointing out errors in the previous version. 
The authors sincerely thank the anonymous referees for suggesting improvements which lead them to find the exact analytical solutions instead of series approximations as in the previous version.

\section{References}


\begin{thebibliography}{88}

\bibitem{Moore:2013sra} 
D.~G.~Moore and V.~H.~Satheeshkumar,
Int.\ J.\ Mod.\ Phys.\ D {\bf 22}, 1342026 (2013)
[arXiv:1305.7221 [gr-qc]].


\bibitem{Bars:2019lek} 
H.~Pihan-Le Bars {\it et al.},
Phys.\ Rev.\ Lett.\  {\bf 123}, no. 23, 231102 (2019)
[arXiv:1912.03030 [physics.space-ph]].


\bibitem{Jacobson:2000xp} 
T.~Jacobson and D.~Mattingly,
Phys.\ Rev.\ D {\bf 64}, 024028 (2001)
[gr-qc/0007031].

\bibitem{Eling:2003rd} 
C.~Eling and T.~Jacobson,
Phys.\ Rev.\ D {\bf 69}, 064005 (2004)
[gr-qc/0310044].

\bibitem{Jacobson:2004ts} 
T.~Jacobson and D.~Mattingly,
Phys.\ Rev.\ D {\bf 70}, 024003 (2004)
[gr-qc/0402005].

\bibitem{Eling:2004dk} 
C.~Eling, T.~Jacobson and D.~Mattingly,
in ``Deserfest: A celebration of the life and works of Stanley Deser. Proceedings, Meeting, Ann Arbor, USA, April 3-5, 2004,''
ed: J.~T.~Liu, M.~J.~Duff, K.~S.~Stelle and R.~P.~Woodard. 
gr-qc/0410001.


\bibitem{Foster:2005dk} 
B.~Z.~Foster and T.~Jacobson,
Phys.\ Rev.\ D {\bf 73}, 064015 (2006)
[gr-qc/0509083].


\bibitem{Eling2006}
C.~Eling and T.~Jacobson,
Class.\ Quant.\ Grav.\  {\bf 23}, 5643 (2006)
Erratum: [Class.\ Quant.\ Grav.\  {\bf 27}, 049802 (2010)]
[gr-qc/0604088].

\bibitem{Eling2007} 
C.~Eling, T.~Jacobson and M.~Coleman Miller,
Phys.\ Rev.\ D {\bf 76}, 042003 (2007)
Erratum: [Phys.\ Rev.\ D {\bf 80}, 129906 (2009)]
[arXiv:0705.1565 [gr-qc]].

\bibitem{Foster2006} 
B.~Z.~Foster,
Phys.\ Rev.\ D {\bf 73}, 024005 (2006)
[gr-qc/0509121].

\bibitem{Garfinkle2007} 
D.~Garfinkle, C.~Eling and T.~Jacobson,
Phys.\ Rev.\ D {\bf 76}, 024003 (2007)
[gr-qc/0703093 [GR-QC]].

\bibitem{Konoplya2007}
R.~A.~Konoplya and A.~Zhidenko,
Phys.\ Lett.\ B {\bf 644}, 186 (2007)
[gr-qc/0605082].

\bibitem{Tamaki2008} 
T.~Tamaki and U.~Miyamoto,
Phys.\ Rev.\ D {\bf 77}, 024026 (2008)
[arXiv:0709.1011 [gr-qc]].

\bibitem{Barausse2011} 
E.~Barausse, T.~Jacobson and T.~P.~Sotiriou,
Phys.\ Rev.\ D {\bf 83}, 124043 (2011)
[arXiv:1104.2889 [gr-qc]].

\bibitem{Gao:2013im} 
C.~Gao and Y.~G.~Shen,
Phys.\ Rev.\ D {\bf 88}, 103508 (2013)
[arXiv:1301.7122 [gr-qc]].

\bibitem{Ding:2015kba} 
C.~Ding, A.~Wang and X.~Wang,
Phys.\ Rev.\ D {\bf 92}, no. 8, 084055 (2015)
[arXiv:1507.06618 [gr-qc]].

\bibitem{Ding2016} 
C.~Ding, C.~Liu, A.~Wang and J.~Jing,
Phys.\ Rev.\ D {\bf 94}, no. 12, 124034 (2016)
[arXiv:1608.00290 [gr-qc]].

\bibitem{Ding2016a}
C.~Ding, A.~Wang, X.~Wang and T.~Zhu,
Nucl.\ Phys.\ B {\bf 913}, 694 (2016)
[arXiv:1512.01900 [gr-qc]].

\bibitem{Barausse2016}
E.~Barausse, T.~P.~Sotiriou and I.~Vega,
Phys.\ Rev.\ D {\bf 93}, no. 4, 044044 (2016)
[arXiv:1512.05894 [gr-qc]].

\bibitem{Latta:2016jix} 
J.~Latta, G.~Leon and A.~Paliathanasis,
JCAP {\bf 1611}, 051 (2016)
[arXiv:1606.08586 [gr-qc]].

\bibitem{Lin2019}
K.~Lin, F.~H.~Ho and W.~L.~Qian, 
Int.\ J.\ Mod.\ Phys.\ D {\bf 28}, no. 03, 1950049 (2018)
[arXiv:1704.06728 [gr-qc]].

\bibitem{Ding:2017gfw} 
C.~Ding,
Phys.\ Rev.\ D {\bf 96}, no. 10, 104021 (2017)
[arXiv:1707.06747 [gr-qc]].

\bibitem{Bhattacharjee:2018nus} 
M.~Bhattacharjee, S.~Mukohyama, M.~B.~Wan and A.~Wang,
Phys.\ Rev.\ D {\bf 98}, no. 6, 064010 (2018)
[arXiv:1806.00142 [gr-qc]].

\bibitem{Lin:2018ken} 
K.~Lin {\it et al.},
Phys.\ Rev.\ D {\bf 99}, no. 2, 023010 (2019)
[arXiv:1810.07707 [astro-ph]].

\bibitem{Zhu2019}
T.~Zhu, Q.~Wu, M.~Jamil and K.~Jusufi,
Phys.\ Rev.\ D {\bf 100}, no. 4, 044055 (2019)
[arXiv:1906.05673 [gr-qc]].

\bibitem{Ding2019}
C.~Ding,
Nucl.\ Phys.\ B {\bf 938}, 736 (2019)
[arXiv:1812.07994 [gr-qc]].

\bibitem{Coley:2019tyx} 
A.~Coley and G.~Leon,
Gen.\ Rel.\ Grav.\  {\bf 51}, no. 9, 115 (2019)
[arXiv:1905.02003 [gr-qc]].

\bibitem{Leon:2019jnu} 
G.~Leon, A.~Coley and A.~Paliathanasis,
Annals Phys.\  {\bf 412}, 168002 (2020)
[arXiv:1906.05749 [gr-qc]].

\bibitem{Zhang:2019iim} 
C.~Zhang, X.~Zhao, A.~Wang, B.~Wang, K.~Yagi, N.~Yunes, W.~Zhao and T.~Zhu,
Phys.\ Rev.\ D {\bf 101}, no. 4, 044002 (2020)
[arXiv:1911.10278 [gr-qc]].

\bibitem{Zhang:2020too}
C.~Zhang, X.~Zhao, K.~Lin, S.~Zhang, W.~Zhao and A.~Wang,
[arXiv:2004.06155 [gr-qc]].


\bibitem{Penrose:1964wq}
R.~Penrose,
Phys. Rev. Lett. \textbf{14}, 57-59 (1965)

\bibitem{Nobel2020}
The Nobel Prize in Physics 2020. NobelPrize.org. Nobel Media AB 2020. Sun. 11 Oct 2020. https://www.nobelprize.org/prizes/physics/2020/summary/ 

\bibitem{Penrose:1969pc}
R.~Penrose,
Riv. Nuovo Cim. \textbf{1}, 252-276 (1969)

\bibitem{rev1} R. Penrose, in {\it Black holes and relativistic stars}, ed. R. M. Wald, University of Chicago Press (1998). 

\bibitem{rev2} 
A.~Krolak,
Prog. Theor. Phys. Suppl. \textbf{136}, 45-56 (1999)
[arXiv:gr-qc/9910108 [gr-qc]].


\bibitem{rev3}
P.~S.~Joshi,
Pramana \textbf{55}, 529-544 (2000)
[arXiv:gr-qc/0006101 [gr-qc]].
%
\bibitem{rev4} 
M.~N.~Celerier and P.~Szekeres,
Phys. Rev. D \textbf{65}, 123516 (2002)
[arXiv:gr-qc/0203094 [gr-qc]].
%
\bibitem{rev5} 
R.~Giambo, F.~Giannoni, G.~Magli and P.~Piccione,
Commun. Math. Phys. \textbf{235}, 545-563 (2003)
[arXiv:gr-qc/0204030 [gr-qc]].

\bibitem{rev6} 
T.~Harada, H.~Iguchi and K.~i.~Nakao,
Prog. Theor. Phys. \textbf{107}, 449-524 (2002)
[arXiv:gr-qc/0204008 [gr-qc]].

\bibitem{rev7} 
C.~F.~C.~Brandt, L.~M.~Lin, J.~F.~Villas da Rocha and A.~Z.~Wang,
Int. J. Mod. Phys. D \textbf{11}, 155-186 (2002)
[arXiv:gr-qc/0105019 [gr-qc]].

\bibitem{rev8} 
R.~Chan, M.~F.~A.~da Silva and J.~F.~Villas da Rocha,
Int. J. Mod. Phys. D \textbf{12}, 347-368 (2003)
[arXiv:gr-qc/0209067 [gr-qc]].

\bibitem{Joshi:2015uoq}
P.~S.~Joshi,
\textit{The Story of Collapsing Stars: Black Holes, Naked Singularities, and the Cosmic Play of Quantum Gravity}, 
Oxford University Press (2015).

\bibitem{Joshi2014}
P.~S.~Joshi, D.~Malafarina and R.~Narayan,
Class. Quant. Grav. \textbf{31}, 015002 (2014)
[arXiv:1304.7331 [gr-qc]].
%
\bibitem{Chakraborty2017a} 
C.~Chakraborty, P.~Kocherlakota and P.~S.~Joshi,
Phys. Rev. D \textbf{95}, no.4, 044006 (2017)
[arXiv:1605.00600 [gr-qc]].
%
\bibitem{Chakraborty2017b} 
C.~Chakraborty, M.~Patil, P.~Kocherlakota, S.~Bhattacharyya, P.~S.~Joshi and A.~Królak,
Phys. Rev. D \textbf{95}, no.8, 084024 (2017)
[arXiv:1611.08808 [gr-qc]].
%
\bibitem{Chakraborty2018} 
C.~Chakraborty and S.~Bhattacharyya,
Phys. Rev. D \textbf{98}, no.4, 043021 (2018)
[arXiv:1712.01156 [astro-ph.HE]].
%
\bibitem{Chakraborty2019} 
C.~Chakraborty and S.~Bhattacharyya,
JCAP \textbf{05}, 034 (2019)
[arXiv:1901.04233 [astro-ph.HE]].
%
\bibitem{Shaikh2019} 
R.~Shaikh and P.~S.~Joshi,
JCAP \textbf{10}, 064 (2019)
[arXiv:1909.10322 [gr-qc]].

\bibitem{Wang} 
K.~Lin, O.~Goldoni, M.~F.~da Silva and A.~Wang,
Phys. Rev. D \textbf{91}, no.2, 024047 (2015)
[arXiv:1410.6678 [gr-qc]].

\bibitem{Berglund2012} 
P.~Berglund, J.~Bhattacharyya and D.~Mattingly,
Phys. Rev. D \textbf{85}, 124019 (2012)
[arXiv:1202.4497 [hep-th]].

\bibitem{Oost2018} 
J.~Oost, S.~Mukohyama and A.~Wang,
Phys. Rev. D \textbf{97}, no.12, 124023 (2018)
[arXiv:1802.04303 [gr-qc]].

\end{thebibliography}
\end{document}